\documentclass[a4paper]{jpconf}
\usepackage[pdftex]{graphicx}
\usepackage{epstopdf}

\def\arcsec{\hbox{$^{\prime\prime}$}\,}
\def\lesssim{\mathrel{\hbox{\rlap{\hbox{\lower4pt\hbox{$\sim$}}}\hbox{$<$}}}}
\def\gtrsim{\mathrel{\hbox{\rlap{\hbox{\lower4pt\hbox{$\sim$}}}\hbox{$>$}}}}
\def\mic{\hbox{\,$\mu$m}\,}

\begin{document}
\title{Nuclear mid-infrared properties of nearby low-luminosity AGN}

\author{D Asmus$^{1}$, S F H\"onig$^{2}$, P Gandhi$^{3}$, A Smette$^{4}$ and W J Duschl$^{1,5}$ }
\address{$^1$ 
Institut f\"ur Theoretische Physik und Astrophysik, Christian- Albrechts-Universit\"at zu Kiel, \hspace*{0.15cm} Leibnizstr. 15, 24098 Kiel Germany              
}

\address{$^2$ Department of Physics, University of California in Santa Barbara, Broida Hall, Santa \hspace*{0.15cm} Barbara, CA 93106-9530, USA}
\address{$^3$ Institute of Space and Astronautical Science (ISAS), Japan Aerospace Exploration Agency, \hspace*{0.15cm}  3-1-1 Yoshinodai, chuo-ku, Sagamihara, Kanagawa 252-5210, Japan}
\address{$^4$ European  Southern Observatory, Casilla 19001, Santiago 19, Chile}
\address{$^5$ Steward Observatory, The University of Arizona, 933 N. Cherry Ave, Tucson, AZ 85721, \hspace*{0.15cm} USA}
\ead{asmus@astrophysik.uni-kiel.de}

\begin{abstract}
We present ground-based high-spatial resolution mid-infrared (MIR) observations of 20 nearby low-luminosity AGN (LLAGN) with VLT/VISIR and the preliminary analysis of a new sample of 10 low-luminosity Seyferts observed with Gemini/Michelle.
LLAGN are of great interest because these objects are the most common among active galaxies, especially in the nearby universe. 
Studying them in great detail makes it possible to investigate the AGN evolution over cosmic timescale. 
Indeed, many LLAGN likely represent the final stage of an AGN's lifetime. 
We show that even at low luminosities and accretion rates nuclear unresolved MIR emission is present in most objects.
Compared to lower spatial resolution \textit{Spitzer}/IRS spectra, the high-resolution MIR photometry exhibits significantly lower fluxes and different PAH emission feature properties in many cases.
By using scaled \textit{Spitzer}/IRS spectra of typical starburst galaxies, we show that the star formation contribution to the 12\mic emission is minor in the central parsecs of most LLAGN.
Therefore, the observed MIR emission in the VISIR and Michelle data is most likely emitted by the AGN itself, which, for higher luminosity AGN, is interpreted as thermal emission from a dusty torus. 
Furthermore, the 12\mic emission of the LLAGN is strongly correlated with the absorption corrected 2-10\,keV luminosity and the MIR--X-ray correlation found previously for AGN is extended to a range from $10^{40}$ to $10^{45}$\,erg/s. 
This correlation is independent of the object type, and in particular the low-luminosity Seyferts observed with Michelle fall exactly on the power-law fit valid for brighter AGN.
In addition, no dependency of the MIR--X-ray ratio on the accretion rate is found.
These results are consistent with the unification model being applicable even in the probed low-luminosity regime. 
\end{abstract}

\section{Introduction}
One of the major discoveries of the last two decades in extragalactic astronomy was that the structure and evolution of the central kiloparsec region of galactic nuclei are strongly connected to the properties and evolution of the central super-massive black hole (SMBH) that presumably resides in the center of each galaxy. 
In particular, the growth of the SMBH seems to be related to major galaxy mergers leading to various phases of nuclear activity (see contribution by Duschl et al.).
These active galactic nuclei (AGN) can be described by the unification model of AGN \cite{antonucci_unified_1993} that contains the SMBH in the nucleus, surrounded by an accretion disk and, further outward, a dusty obscuring torus-like structure. 
This torus presumably blocks the direct view onto the central region depending on the viewing angle of the observer, therefore explaining the different AGN sub-classes observed. 
However, for a more accurate description, various modifications to this standard model may arguably be necessary (e.g., as suggested in the contributions by Elvis et al. and Ramos Almeida et al.).
In addition, in the low-luminosity regime various changes in the structure of the AGN might occur.
This is indicated by a different spectral energy distribution (SED) observed in these low-luminosity AGN (LLAGN)\cite{ho_nuclear_2008}, theoretically explained by a change in the accretion disk structure from an optically-thick and geometrically-thin accretion disk \cite{shakura_black_1973} to an optically-thin and geometrically-thick accretion flow (e.g., advection-dominated accretion flow; ADAF \cite{narayan_advection-dominated_1994}) at low accretion rates (Eddington ratio $\eta_\mathrm{Edd} \lesssim 10^{-2}$).
At even lower rates ($\eta_\mathrm{Edd} \lesssim 10^{-5}$), a jet of outflowing plasma along the rotation axis of the accretion disk might even dominate the whole SED \cite{falcke_jet_2000}.
Finally, theoretical work predicts that the dusty obscuring torus will disappear at low luminosities/accretion rates \cite{hoenig_active_2007, elitzur_agn-obscuring_2006}. Observational indications for its disappearance have been found mainly for radio-loud LLAGN \cite{donato_obscuration_2004}, while remaining unproven for the LLAGN in general (but see also  contribution by Fern\'andez Ontiveros et al.).

Searches for the putative torus are most promising in the mid-infrared (MIR), where the thermal emission of the dust is peaking. 
In particular, ground-based high-spatial resolutions MIR observations yield valuable information about this probably compact ($\lesssim 1$\,pc) structure.
It was found that the observed MIR emission of local AGN can be best explained by a torus consisting of many clouds \cite{hoenig_dusty_2010-1,alonso-herrero_torus_2011} (and also contribution by Ramos Almeida et al.).  
Although MIR interferometry provides an excellent possibility to investigate and resolve the torus in nearby AGN (see contributions by Burtscher et al., Kishimoto et al., and  Tristram et al.), current its sensitivity is not sufficient to detect most LLAGN. 

Even though lacking the high-spatial resolution, early MIR observations led to the discovery of a strong correlation between the nuclear MIR and X-ray emission \cite{krabbe_n-band_2001, lutz_relation_2004}.
This MIR--X-ray correlation can in principle be understood by a close connection between the ultraviolet and X-ray emission originating in the inner accretion disk region. 
While the hard X-ray emission can penetrate the obscuring torus almost unabsorbed as long as the column density remains in the Compton-thin regime, the ultraviolet is absorbed by the dust and re-emitted in the MIR as thermal radiation. 
Note that this explanation is independent of the torus orientation, because the MIR emission is nearly isotropic.
Therefore hard X-rays are also a good direct tracer of the accretion activity and the current intrinsic power output of an AGN. 
By using high-spatial resolution MIR VLT/VISIR imaging, we could recently demonstrate for a sample of $\sim 40$ local AGN that the MIR--X-ray correlation is indeed very strong, in particular when the hard 2-10\,keV X-ray emission is corrected for absorption \cite{gandhi_resolving_2009}.    
The present work represents a direct extension of our previous work into the low-luminosity regime and is driven by the main goal of probing the existence of a torus-like obscuring structure in LLAGN. 

\begin{figure}[th]

 \includegraphics[width=0.7\linewidth]{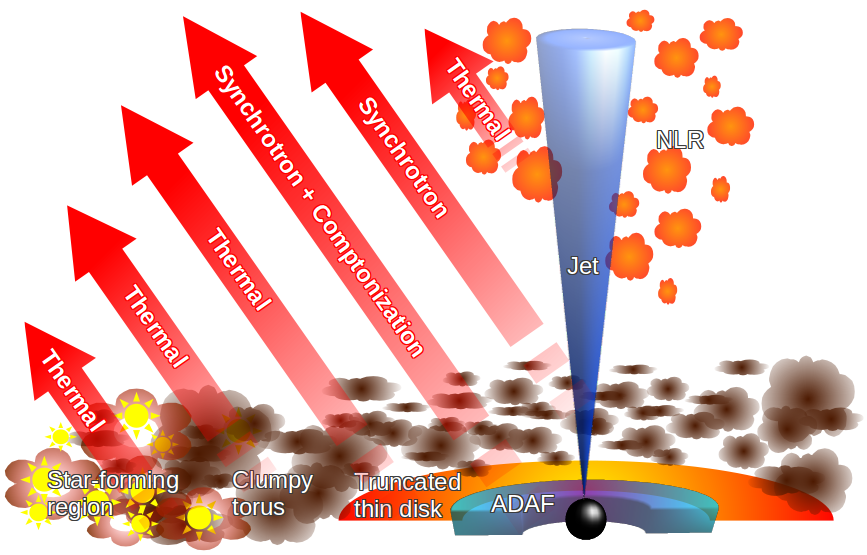}
\hspace{0.05\linewidth}%
\begin{minipage}[b]{0.25\linewidth}
 \caption{\label{fig:1}
Schematic view of the central structure of an LLAGN. The red arrows indicate the MIR emission of the various components and its nature.
}
\end{minipage}
\end{figure}

However, because of the large number of possible MIR emitters in this luminosity regime (Fig.\ref{fig:1}), only a first step will be presented: high-spatial resolution Gemini/Michelle and VLT/VISIR MIR imaging of LLAGN.
We can investigate the validity of the MIR--X-ray correlation for LLAGN and constrain possible contamination by circum-nuclear star formation affecting the MIR photometry. 
Star formation can significantly contribute or even dominate the emission at low luminosities. 
For a detailed description of the VLT/VISIR data we refer to \cite{asmus_mid-infrared_2011-1}, while comparable observations are also presented in the contribution by Mason et al.

\section{Methods}
In order to separate the nuclear MIR emission in LLAGN from surrounding emission regions, high angular resolution data is necessary.
Only this guarantees that non-nuclear contamination is mostly avoided, as demonstrated, e.g., in the contribution by Prieto et al. 
Therefore, we used VLT/VISIR and Gemini/Michelle imaging in narrow-band filters to probe the 12\mic continuum and polycyclic aromatic hydrocarbon (PAH) at 11.3\,$\mu$m for two samples of nearby LLAGN.
These were selected to have absorption-corrected X-ray fluxes $F_{2-10\mathrm{keV}} > 10^{-12}$erg/s/cm$^2$, taken from the literature, mainly  \cite{gonzalez-martin_x-ray_2009} and \cite{panessa_x-ray_2006}.
The X-ray flux limit corresponds to an estimated 12\mic flux of 10\,mJy, assuming the MIR--X-ray correlation from \cite{gandhi_resolving_2009}, and represents roughly the lowest detectable flux with the ground-based MIR instruments that we used.
The sample consists of LLAGN with a mean X-ray luminosity $\langle \log L_{2-10\mathrm{keV}}/\mathrm{erg\,s^{-1}} \rangle = 40.4$ at a mean redshift of  $\langle z \rangle = 0.005$, corresponding to a distance of $\langle D \rangle = 19$\,Mpc and a resolved physical scale of $\langle r_0 \rangle = 40$\,pc.
The southern sample of 18 LLAGN was observed in 2009, between April and September, with VLT/VISIR in the PAH2 ($11.25 \pm 0.59\mic$) and NeIIref1 ($12.27 \pm 0.18\mic$) filters, and is described in more detail in \cite{asmus_mid-infrared_2011-1}.
These LLAGN have mixed properties, e.g., optical classification and are not complete in any sense.
The northern sample consists of 10 low-luminosity Seyferts drawn from the Palomar LLAGN survey \cite{ho_search_1997-1} and were observed with Gemini/Michelle in February 2010 in the Si-5 ($11.6 \pm 0.55\mic$) and Si-6 ($12.5 \pm 0.6\mic$) filters.
All observations were performed in the standard chopping and nodding mode.
The data were reduced with the observatory delivered pipeline packages and photometric fluxes were measured with custom IDL routines \cite{hoenig_dusty_2010-1}. 

\section{Results}
We detected 7 out of 18 southern and all 10 northern LLAGN with all detected objects being visible in both filters. 
The high number of non-detected LLAGN can be explained by the high uncertainties in the X-ray luminosities that were used for the sample selection.
Indeed, many of these have been revised to much lower values in more recent X-ray observations (further discussed in \cite{asmus_mid-infrared_2011-1}).  
Therefore we included 9 additional LLAGN with VISIR observations taken from the literature.
All detected LLAGN appear point-like in the MIR at an angular resolution of $\sim 0.4\arcsec$, and no extended or off-nuclear emission is visible in the central 4\arcsec around these nuclei, similar to the findings for most of the brighter AGN \cite{horst_mid-infrared_2009}.
The corresponding VISIR images can be found in \cite{asmus_mid-infrared_2011-1} and a detailed analysis of the Michelle data will be presented in a future work.

\subsection{Comparison to \textit{Spitzer} -- star formation contribution}
\begin{figure}[th]
 \includegraphics[width=0.65\linewidth]{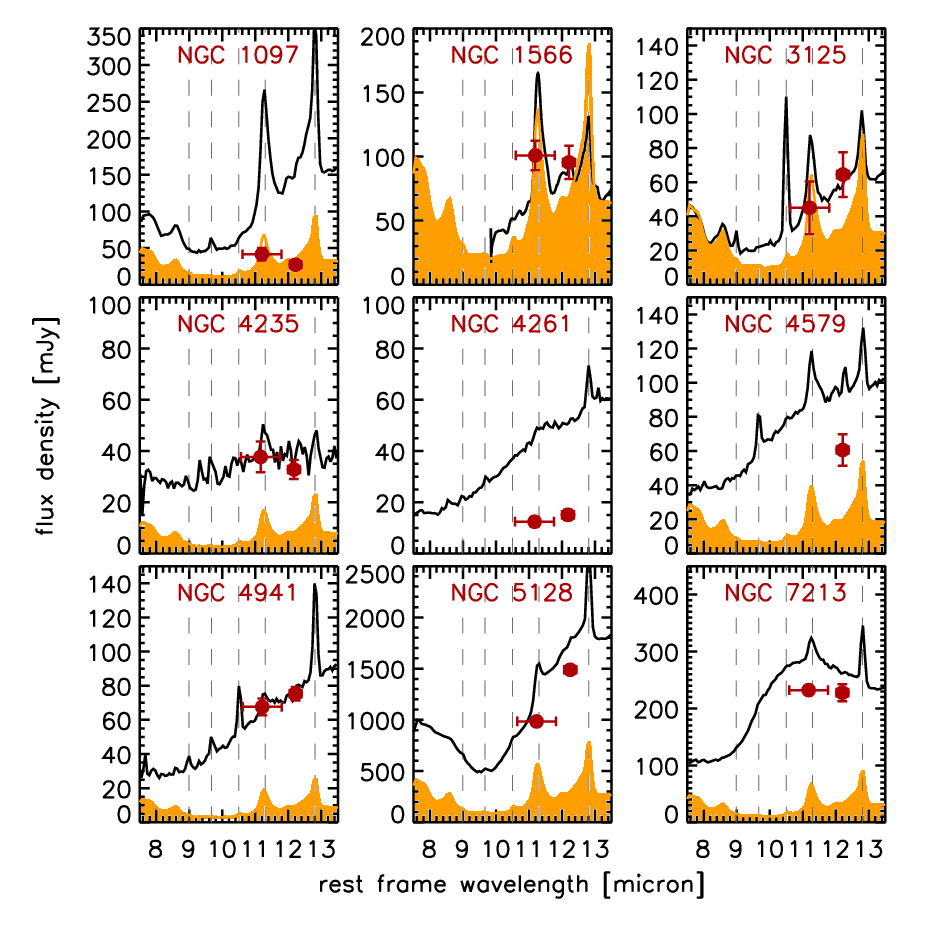}
\hspace{0.05\linewidth}%
\begin{minipage}[b]{0.3\linewidth}
 \caption{\label{fig:2}
Comparison of VLT/VISIR photometry (red symbols) and {\it Spitzer} IRS spectra (black line) for LLAGN. 
Horizontal error bars correspond to the filter pass band.
Commonly observed emission lines are indicated by the dotted lines. 
In addition, the scaled starburst template SED (orange-filled curves) correspond to the derived upper limit for the starburst contribution to the VISIR fluxes.
}
\end{minipage}
\end{figure}

Because \textit{Spitzer}/IRS spectra (angular resolution $\sim 4\arcsec$) are widely used throughout the literature to infer AGN properties, a comparison with the Michelle and VISIR photometry is of great interest. 
In Fig.~\ref{fig:2} we show a selection of objects from the VISIR sample.
While the fluxes agree well for some objects, e.g., NGC 1566, the majority of objects ($\sim 60\%$) show significantly lower fluxes in VISIR than in \textit{Spitzer} data (e.g, NGC 1097, NGC 4261, NGC 4579). 
In particular, many LLAGN do not possess PAH 11.3\mic emission on a 0.4\arcsec scale (e.g. NGC 7213). 
However, NGC 1097 and NGC 1566 most likely exhibit nuclear PAH 11.3\mic emission as indicated by the comparably high PAH2 filter fluxes.
We attribute the missing flux to extended diffuse emission regions that surround the nucleus on a 4\arcsec scale. 
These have a low surface brightness and, therefore, are over-resolved and not visible in the VISIR and Michelle images while being included in the \textit{Spitzer} beam.
In general, there is no evidence for either any correlation between the VISIR and \textit{Spitzer} fluxes, or any trend with the optical classification. 

PAH emission is usually attributed to star formation and hence routinely used as its tracer. 
Furthermore, a tight correlation between the PAH 11.3\mic emission line flux and the 12\mic continuum is present in local starburst galaxies \cite{asmus_mid-infrared_2011-1}.
Reversely we use the PAH emission to estimate the amount of MIR continuum emission due to star formation in the innermost 0.4\arcsec of the observed LLAGN.  
For this purpose we construct a starburst template SED by normalizing the IRS spectra from \cite{brandl_mid-infrared_2006} by their PAH 11.3\mic flux.
This template SED can in turn be applied to the individual LLAGN by scaling with the PAH 11.3\mic emission occurring in those objects.
With the data at hand, this PAH emission can only be constrained by using either the IRS spectrum or the filter containing this feature (i.e. PAH2 for VISIR), depending on which provides the lower upper-limit.
Hence, the scaled starburst template SEDs (orange-filled curves in Fig.~\ref{fig:2}) in turn represent upper limits for the star formation related MIR emission. 
For the majority of LLAGN ($\sim 70\%$), we can constrain the 12\mic continuum flux contamination through star formation to be less than $50\%$ on a 0.4\arcsec scale, while for the remaining $30\%$ the contamination can possibly be up to $100\%$.  

\subsection{MIR--X-ray correlation}
\begin{figure}[th]
\includegraphics[width=0.6\linewidth]{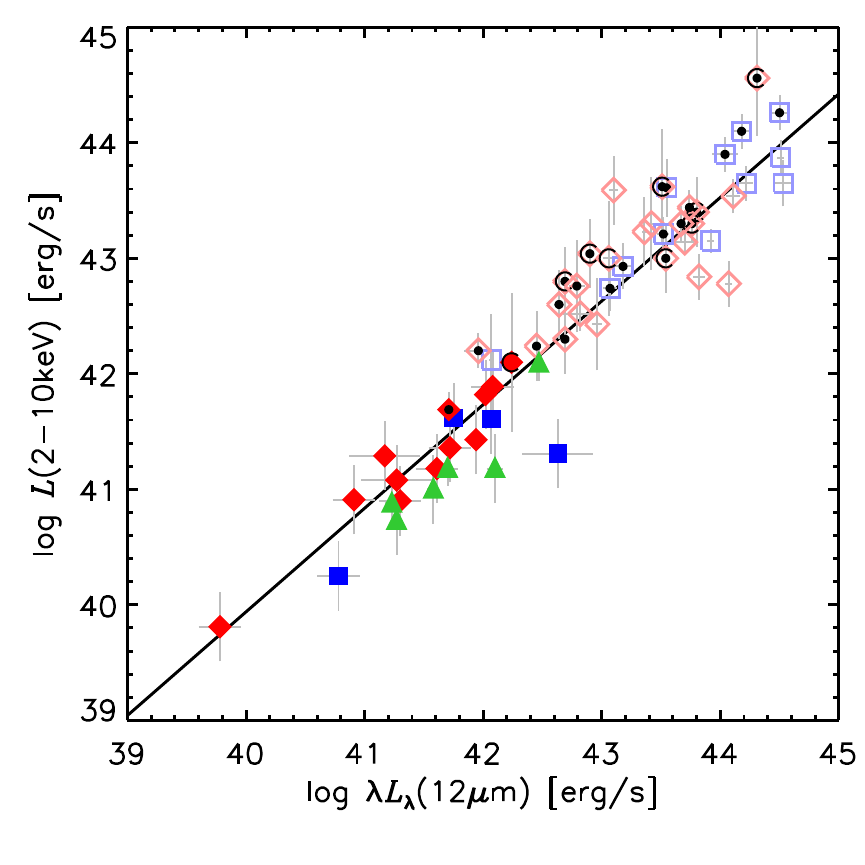}
\hspace{0.05\linewidth}%
\begin{minipage}[b]{0.35\linewidth}
\caption{\label{fig:3}
Absorption-corrected hard X-ray luminosities vs. the nuclear MIR luminosities for LLAGN (preliminary; filled symbols) and the AGN from \cite{gandhi_resolving_2009} (empty symbols);
blue squares: type 1 Seyferts (type 1.5 or lower);
red diamonds: type 2 Seyferts; 
green triangles: LINERs;
objects highlighted with central black-filled circles: ``well-resolved'' AGN from \cite{gandhi_resolving_2009} ;
solid line: power-law fit to all LLAGN and AGN.               
}
\end{minipage}
\end{figure}

As mentioned in Section~1, a strong correlation exists between the absorption-corrected hard X-ray and the observed 12\mic emission for local AGN.
Fig.~\ref{fig:3} shows the luminosities at these wavelengths for the LLAGN and the sample of local AGN from \cite{gandhi_resolving_2009}. 
The LLAGN tightly follow the same correlation  and extend it down to luminosities of $\sim 10^{40}$\,erg/s. 
However, the observed LINERs might exhibit a minor systematic offset ($\sim 0.3$\,dex) towards higher MIR or lower X-ray luminosities. 
Nevertheless, the numbers are too low for any statistically significant difference, and in general, the correlation remains independent of the optical classification within the measurement uncertainties.
A power-law fit using \texttt{fitexy} \cite{press_numerical_1992} yields:
\begin{equation}
   \mathrm{log} \left( \frac{\lambda L_\lambda(12\mic) } {10^{43}\,\mathrm{erg\,s}^{-1}} \right) = ( 0.41 \pm 0.03) + ( 1.12 \pm 0.04 )  \mathrm{ log}\left( \frac{L_\mathrm{2-10keV} } {10^{43}\,\mathrm{erg\,s}^{-1} } \right),
\end{equation}
which is similar to previous results \cite{gandhi_resolving_2009}.

\subsection{Dependency on accretion rate}
One of the fundamental parameters determining the AGN structure is assumed to be the accretion rate, which can be estimated by the Eddington ratio  $\eta_\mathrm{Edd} = \log L_\mathrm{bol} / L_\mathrm{Edd}$, with $L_\mathrm{bol}$ being the bolometric luminosity of the AGN, $L_\mathrm{Edd} = 1.26 \times 10^{38} (M_\mathrm{BH} /M_\odot) $\,erg/s its Eddington luminosity depending on $M_\mathrm{BH}$, the mass of the black hole.
We estimate the black hole masses of the LLAGN mainly by using the $M_\mathrm{BH}$-$\sigma$ correlation \cite{gueltekin_m-_2009} with the stellar velocity dispersions $\sigma$ taken from the literature (mainly \cite{ho_search_2009} and the Hyperleda database \cite{paturel_hyperleda._2003}).   
Furthermore, the bolometric luminosity is simply assumed as $L_\mathrm{bol} = 10 L_\mathrm{2-10keV}$, which seems reasonable for LLAGN \cite{ho_nuclear_2008}. 

The Eddington ratios of all observed LLAGN and the AGN from \cite{gandhi_resolving_2009} extend over a large range of 5 orders of magnitude, $10^{-5} \lesssim \eta_\mathrm{Edd} \lesssim 1 $.
The MIR--X-ray luminosity ratio is constant, $ \log (\lambda L_\lambda(12\mic)/L_\mathrm{2-10keV}) = 0.31 \pm 0.36$, over the whole range of $\eta_\mathrm{Edd}$, and therefore independent of the accretion rate for all observed AGN.

\section{Discussion \& conclusions}
The high-angular resolution MIR imaging of $\sim 30$ nearby LLAGN showed point-like nuclear emission at $\sim 50$\,pc scale in all detected objects. 
No correlation to larger aperture data like Spitzer is evident from a comparison of VISIR photometry and IRS spectra.
Therefore one has to be careful when deriving AGN properties from these larger aperture data. 
In many cases the observed fluxes are significantly lower and spectral features, as the PAH 11.3\mic emission feature are absent on smaller scales.
Furthermore, by using a starburst template SED, we showed that only $\lesssim 30\%$ of the high-resolution data may still be dominated by MIR emission from star formation occurring on $\lesssim 50$\,pc scale.
The majority ($\gtrsim 70\%$) of LLAGN are, however, dominated by AGN emission on this scale, while its exact origin -- torus, jet or accretion disk -- remains undetermined at this point.    
Furthermore, the MIR--X-ray correlation is valid for all AGN observed at high-angular resolution in the MIR over the whole probed luminosity range from $10^{40}$ to $\sim 10^{45}$\,erg/s. 
Owing to the fact that no dependence of either optical classification, or radio-loudness, or host morphology was found, this correlation might put strong constraints on the universal AGN structure.  
Interestingly, none of the theoretically predicted changes in the accretion structure with decreasing accretion rate, e.g., onset of ADAF or the jet-dominance, seem to have any imprint on the MIR--X-ray luminosity ratio. 
However, the sample is likely incomplete (selection based on availability of 2-10\,keV data) and possibly biased against highly obscured (Compton-thick) objects.
This along with the small total sample size prevents us from reaching any final conclusions on the AGN structure and nature of the MIR emitter in the low-luminosity regime.
At this point we can only conclude that the results agree with those for brighter AGN, for which the MIR emission is well explained by a dusty clumpy torus.
Therefore, such a torus might also exist in LLAGN, i.e. the standard unification model of AGN might still hold in this regime, in agreement with \cite{panessa_x-ray_2006,maoz_low-luminosity_2007}.
Furthermore, these results imply that in LINERs, the emission is AGN-dominated in at least central tens of parsecs (compare \cite{yan_nature_2011}).     
Finally, the MIR--X-ray correlation represents a powerful observational tool to convert between both wavelength bands for any AGN. 
In order to settle the question about the nature of the MIR emission, SED modeling with particular focus on the high-resolution infrared data seems most promising (see also contributions from Prieto et al. and Fern\'andez Ontiveros et al.).

\section*{References}
\bibliographystyle{iopart-num} 
\bibliography{Asmus_ref.bib} 

\providecommand{\newblock}{}
\begin{thebibliography}{10}
\expandafter\ifx\csname url\endcsname\relax
  \def\url#1{{\tt #1}}\fi
\expandafter\ifx\csname urlprefix\endcsname\relax\def\urlprefix{URL }\fi
\providecommand{\eprint}[2][]{\url{#2}}

\bibitem{antonucci_unified_1993}
Antonucci R 1993 {\em ARA\&A\/} {\bf 31} 473--521

\bibitem{ho_nuclear_2008}
Ho L~C 2008 {\em ARA\&A\/} {\bf 46} 475--539

\bibitem{shakura_black_1973}
Shakura N~I and Sunyaev R~A 1973 {\em A\&A\/} {\bf 24} 337--355

\bibitem{narayan_advection-dominated_1994}
Narayan R and Yi I 1994 {\em ApJ\/} {\bf 428} L13--L16

\bibitem{falcke_jet_2000}
Falcke H and Markoff S 2000 {\em A\&A\/} {\bf 362} 113--118

\bibitem{hoenig_active_2007}
H\"onig S~F and Beckert T 2007 {\em MNRAS\/} {\bf 380} 1172--1176

\bibitem{elitzur_agn-obscuring_2006}
Elitzur M and Shlosman I 2006 {\em ApJ\/} {\bf 648} L101--L104

\bibitem{donato_obscuration_2004}
Donato D, Sambruna R~M and Gliozzi M 2004 {\em ApJ\/} {\bf 617} 915--929

\bibitem{hoenig_dusty_2010-1}
H\"onig S~F, Kishimoto M, Gandhi P, Smette A, Asmus D, Duschl W, Polletta M and
  Weigelt G 2010 {\em A\&A\/} {\bf 515} 23

\bibitem{alonso-herrero_torus_2011}
{Alonso-Herrero} A, {Ramos Almeida} C, Mason R, Asensio~Ramos A, Roche P~F,
  Levenson N~A, Elitzur M, Packham C, Rodr\'iguez~Espinosa J~M, Young S,
  {D\'iaz-Santos} T and {P\'erez-Garc\'ia} A~M 2011 {\em ApJ\/} {\bf 736} 82

\bibitem{krabbe_n-band_2001}
Krabbe A, B\"oker T and Maiolino R 2001 {\em ApJ\/} {\bf 557} 626--636

\bibitem{lutz_relation_2004}
Lutz D, Maiolino R, Spoon H~W~W and Moorwood A~F~M 2004 {\em A\&A\/} {\bf 418}
  465--473

\bibitem{gandhi_resolving_2009}
Gandhi P, Horst H, Smette A, H\"onig S, Comastri A, Gilli R, Vignali C and
  Duschl W 2009 {\em A\&A\/} {\bf 502} 457--472

\bibitem{asmus_mid-infrared_2011-1}
Asmus D, Gandhi P, Smette A, H\"onig S~F and Duschl W~J 2011 {\em A\&A\/} {\bf
  536} 36

\bibitem{gonzalez-martin_x-ray_2009}
{Gonz\'alez-Mart\'in} O, Masegosa J, M\'arquez I, Guainazzi M and
  {Jim\'enez-Bail\'on} E 2009 {\em A\&A\/} {\bf 506} 1107--1121

\bibitem{panessa_x-ray_2006}
Panessa F, Bassani L, Cappi M, Dadina M, Barcons X, Carrera F~J, Ho L~C and
  Iwasawa K 2006 {\em A\&A\/} {\bf 455} 173--185

\bibitem{ho_search_1997-1}
Ho L~C, Filippenko A~V and Sargent W~L~W 1997 {\em ApJSS\/} {\bf 112} 315

\bibitem{horst_mid-infrared_2009}
Horst H, Duschl W~J, Gandhi P and Smette A 2009 {\em A\&A\/} {\bf 495} 137--146

\bibitem{brandl_mid-infrared_2006}
Brandl B~R, {Bernard-Salas} J, Spoon H~W~W, Devost D, Sloan G~C, Guilles S, Wu
  Y, Houck J~R, Weedman D~W, Armus L, Appleton P~N, Soifer B~T, Charmandaris V,
  Hao L, Higdon J~A~M and Herter T~L 2006 {\em ApJ\/} {\bf 653} 1129--1144

\bibitem{press_numerical_1992}
Press W~H, Teukolsky S~A, Vetterling W~T and Flannery B~P 1992 {\em Numerical
  recipes in {FORTRAN.} The art of scientific computing\/}

\bibitem{gueltekin_m-_2009}
G\"ultekin K, Richstone D~O, Gebhardt K, Lauer T~R, Tremaine S, Aller M~C,
  Bender R, Dressler A, Faber S~M, Filippenko A~V, Green R, Ho L~C, Kormendy J,
  Magorrian J, Pinkney J and Siopis C 2009 {\em ApJ\/} {\bf 698} 198--221

\bibitem{ho_search_2009}
Ho L~C, Greene J~E, Filippenko A~V and Sargent W~L~W 2009 {\em ApJSS\/} {\bf
  183} 1--16

\bibitem{paturel_hyperleda._2003}
Paturel G, Petit C, Prugniel P, Theureau G, Rousseau J, Brouty M, Dubois P and
  Cambr\'esy L 2003 {\em A\&A\/} {\bf 412} 45--55

\bibitem{maoz_low-luminosity_2007}
Maoz D 2007 {\em MNRAS\/} {\bf 377} 1696--1710

\bibitem{yan_nature_2011}
{Yan} R and {Blanton} M~R 2011 {\em ArXiv e-prints\/} (\textit{Preprint}
  \eprint{1109.1280})

\end{thebibliography}

\end{document}